\documentstyle{mn}

\input epsf
\def\plotone#1{\centering \leavevmode
\epsfxsize=\columnwidth \epsfbox{#1}}

\newcommand{\op}{Ly$\alpha$\ }
\newcommand{\kms}{\, {\rm km \, s}^{-1} }
\newcommand{\arcm}{\, {\rm arcmin}}

\newcommand{\keV}{\, {\rm keV} }

\newcommand{\K}{\, {\rm K} }
\newcommand{\g}{\, {\rm g} }
\newcommand{\cm}{{\, \rm cm} }
\newcommand{\sr}{{\, \rm sr} }
\newcommand{\yr}{{\, \rm yr} }

\newcommand{\erg}{{\, \rm erg} }
\newcommand{\Mpc}{\, {\rm Mpc} }
\newcommand{\phot}{\, {\rm photons} }

\newcommand{\ovii}{\mbox{O{\scriptsize VII}}}
\newcommand{\oviii}{\mbox{O{\scriptsize VIII}}}

\def\apj{ApJ}                 
\def\apjl{ApJ}

\def\aap{A\&A}                
          
\def\aaps{A\&AS}

\title[Resonant scattering by IGM]{Resonant scattering of X-rays by the warm intergalactic medium}

\author[Churazov, Haehnelt, Kotov \& Sunyaev]{E.~Churazov,$^{1,2}$
M.~Haehnelt,$^{1,3}$ 
O.~Kotov,$^{2,1}$ R.~Sunyaev,$^{1,2}$ \\
$^1$ MPI fur Astrophysik, Karl-Schwarzschild-Strasse 1, 85740
Garching, Germany \\
$^2$ Space Research Institute (IKI), Profsouznaya 84/32, Moscow 117810, 
Russia\\
$^3$ Institut for Theoretical Physics, University of California, Santa
Barbara, CA 93106-4030}

\date{Accepted ????????????
      Received ????????????;
      in original form ????????????}

\pagerange{\pageref{firstpage}--\pageref{lastpage}}
\pubyear{2000}

\begin{document}
\maketitle

\label{firstpage}
\begin{abstract}
For the low density filamentary and sheet-like structures in the warm
($\sim 10^4$ to $\sim 10^6$ K) IGM predicted by numerical simulations
the resonant line scattering of X--ray background (XRB) photons by  He
and H--like ions of heavy elements can exceed the ``local'' thermal
emission by a factor of a few or more. Due to the conservative nature
of scattering this resonantly scattered radiation can only be
identified  if a significant fraction of the XRB is resolved and
removed. While the combined spectrum of the resolved sources will
contain X--ray absorption features, the residual background will
contain corresponding emission features with the same intensity. 
At the relevant  densities and temperatures  the lines of  He and H--like
oxygen at 0.57 and 0.65 keV are most promising. 
These lines (which have a typical width of $\sim$ 1--2 eV) may contain
up to 50\% of the total 0.5--1 keV emission of the filament. 
For a nearby ($z \la 0.1$) filament
with a Thomson optical depth of $10^{-4}$  XMM should detect
about 200 photons in the OVII line during a $10^5$ s exposure
if the metallicity of the gas is as large as observed in galaxy clusters.
On average up to a few percent of  the soft XRB  could be resonantly
scattered  by this phase of the IGM  and resonantly scattered photons
should account for a significant fraction of the truly diffuse
background at low energies.
Close to bright  X-ray sources like galaxy  clusters or AGN the flux 
of scattered radiation will be further enhanced. 
Off-line blazars are the most promising illuminating sources. The
scattered emission from AGN may also constrain the duration of the
active phase of these objects.  
\end{abstract}

\section{Introduction}
At high redshift the majority of baryons are  contained in a
photoionized IGM with temperature $\sim 10^4\K $  which is 
responsible for the strong  \op absorption in the spectra of 
high redshift  objects (see Rauch 1998 and Weinberg 1999 for reviews). 
At low redshift, however, there are little observational 
constraints on  the thermal state and baryon content of the IGM
(e.g. Barcons, Fabian \& Rees 1991). Numerical simulations 
predict that a considerable 
fraction of all baryons is still contained in a warm IGM which traces the
filamentary and sheet-like distribution of the dark
matter (e.g. Ostriker \& Cen, 1996, Cen \& Ostriker 1999). 
This phase of the IGM should have densities of a few up to a 
few tens times the  mean baryonic density. The temperatures should
range between a `` photoionization temperature'' at low densities  
($\sim 10^3 -10^4\K$) and the virial temperature of sheets and filaments
(~$\sim 10^6\K$). These temperatures might be enhanced 
due to the energy  input by star formation and AGN and it has  
been argued that such an energy input is  necessary to explain  the 
X-ray luminosity temperature relation  of galaxy clusters
(Kaiser 1991; Ponman, Cannon \& Navarro 1999; Pen 1999). 
This ``warm'' phase of the IGM is of special interest
as it will contain a record of the energy and metals expelled 
from galaxies. The typical surface brightness  of filaments due 
to local thermal emission of the diffuse gas is, however, a factor 
100 or more smaller than that  of the X-ray background. This local thermal 
emission will be detectable  with XMM only for especially 
strong filaments (Pierre, Bryan \& Gastaud 1999) but see Scharf et al. (1999) 
for a claimed detection of such a filament with ROSAT.  
It was also suggested that the warm IGM  produces measurable 
absorption in the resonant transitions of heavy elements 
such as oxygen or iron if there is a bright quasar  behind the gas
(Shapiro and Bahcall 1980, Aldcroft et al., 1994, Hellsten, Gnedin,
Miralda--Escude 1998, Perna and Loeb, 1998, Markevitch, 1999). This
seems a promising method to study  the warm IGM, 
especially as long as  high energy resolution is only possible for
bright sources (i.e. using gratings). It will, however, only give information 
along the line-of-sight  to point  sources.  
We explore here  the  possibility to investigate 
the warm IGM by its emission due to resonant scattering of X-ray 
background photons  by He and H--like ions of heavy elements. 
While for the  high density and  temperatures in galaxy clusters 
the local thermal emission   clearly dominates over resonant 
scattering the opposite is true for the more moderate densities 
and temperatures expected in the sheet-like and filamentary structures
of the warm phase of the IGM. 

In section 2 we discuss the importance of resonant scattering
relative to the local thermal emission due to 
collisional excitation and ionization for the temperatures 
and densities prevalent in the warm IGM. 
Section 3 discusses the detectability of filamentary structures with upcoming
satellite missions. In section 4 we discuss the illumination of
filamentary structures by individual bright sources and section 5
contains our conclusions.  

A Hubble constant of $H_0=50~\kms~\Mpc^{-1}$ 
and an Einstein-de-Sitter Universe was assumed throughout
the paper. For the baryon density we take $\rho_{\rm bar}=
3.6~10^{-31}~\g~\cm^{-3}$ 
($\Omega_{\rm bar} h^2 = 0.02$,  Burles et al., 1999).
The abundances of heavy elements were assumed to be a constant 
fraction of  solar abundances as given by Feldman (1992). 
For the local  XRB spectrum we use the simple approximation $I(E)=I_0\,
E^{-1.3} e^{-\frac{E}{40\keV}}$ for  energies $E \ge 1 \keV$ and a
power law with a photon index of 2 below 1 keV, where  $I_0 =
8~\phot~\cm^{-2}~\sec^{-1}~\keV^{-1}~\sr^{-1}$ (e.g. Barcons \&
Fabian 1992, Miyaji et al. 1998).

\section{Photoionization balance and emissivity of the warm gas}

\subsection{Typical densities and temperatures}
We discuss here the emission from typical sheet-like and filamentary
structures in the warm intergalactic medium taking into account 
the resonant scattering of X-ray background photons. The characteristic 
sheets and filaments seen in numerical simulations are the result
of the non-linear collapse of density perturbation imprinted 
onto the matter distribution in the early Universe. Sheets 
form from  perturbations which collapse
along one axis (Zeldovich 1970) while filaments have collapsed along
two axis.  
This results in typical overdensities of a few in sheets
and a few tens in  filaments.  The typical  Thomson optical depth 
scales linearly with the length scale of the density perturbation and will 
be about $(0.1,0.5) h^{-1} \times10^{-4} (R/8 h^{-1}\Mpc)$ for 
sheets and filaments, respectively. Note that $R= 8 h^{-1}\Mpc$ is 
approximately the scale on which the present-day density field 
has gone  non-linear. The space density of structures 
larger than this decreases exponentially. Typical temperatures 
in the warm IGM  are somewhat uncertain but will be strongly
correlated with density. In the absence of energy input from star 
formation and  AGN  the temperatures at low densities are 
set by the balance of photoheating and adiabatic cooling and 
lie in the range 3000 -10000 K (e.g. Hui \& Gnedin 1997). At higher
densities the gas will be shock-heated. In virialized regions the 
temperatures will be  set by the virial temperature of sheets and  
filaments and should be about $10^6\K$. The energy input 
from star formation and  AGN  will raise this 
temperatures especially at low densities. Ponman, Cannon \& Navarro
(1999) e.g. suggest a minimum ``entropy'' of  $100 h^{-1/3}\keV \cm^2$ 
to explain the X-ray luminosity temperature relation of galaxy clusters 
at the faint end.  

In the discussion below we consider two examples: emission from regions with 
an overdensity of (5,30), corresponding to a electron density of
($10^{-6}~cm^{-3}$,$6~10^{-6}~cm^{-3}$), a temperature of ($2 \times
10^5K$,$10^{6}K$),  and a metallicity of (10\%,30\%) solar which may
resemble a typical sheet and filament, respectively. For a size of 8 Mpc a
sheet and filament will have a Thomson optical depth of $(0.2,1)\times
10^{-4}$ respectively. The emission spectra are calculated below for zero
redshift.

\subsection{Thermal emission of the gas in collisional equilibrium}
For  gas with  given density and temperature in pure collisional
equilibrium (coronal approximation) the following processes 
contribute to the X--ray emission: continuum emission (free-free and
bound-free), recombination lines and emission lines excited by
electron collisions. In the following we call these emission mechanisms
collectively  ``thermal'' emission. We used the code MEKA
\cite{mgo85,mlo86,kaa92} as implemented in the software package
XSPEC v10 \cite{ar96} to calculate this thermal emission. For our
adopted parameters the X--ray emissivity of the gas in filaments and
sheets is extremely low (due to both the low temperature and  the low density
of the gas).   Detecting these structures requires very high
sensitivity. The thermal emission from a typical
filament in the 0.5--1 keV energy band (in  units of
$\keV ~\sec^{-1}~\cm^{-2}~\keV^{-1}~\sr^{-1}$ to facilitate comparison with
the XRB surface brightness) is shown in  Fig. 1 by the dotted
curve. For comparison the intensity of the XRB is shown by the solid
line. For our canonical sheet the emissivity is  several orders
of magnitude lower and  well below the limits of the plot. 

The intrinsic width of the lines should be dominated by the Hubble
flow, peculiar and turbulent velocities rather than thermal
broadening. 
In the considered energy range of 0.5--1 keV the velocity dispersion of
500-1000 $\kms$ implies a width of $\sim 1$--2 eV.
The spectra shown in all figures are convolved  with a Gaussian with a FWHM
of 2 eV. We note here that this kind of resolution may be achieved by
projected X-ray missions  like Constellation-X or XEUS. 

\begin{figure}
\plotone{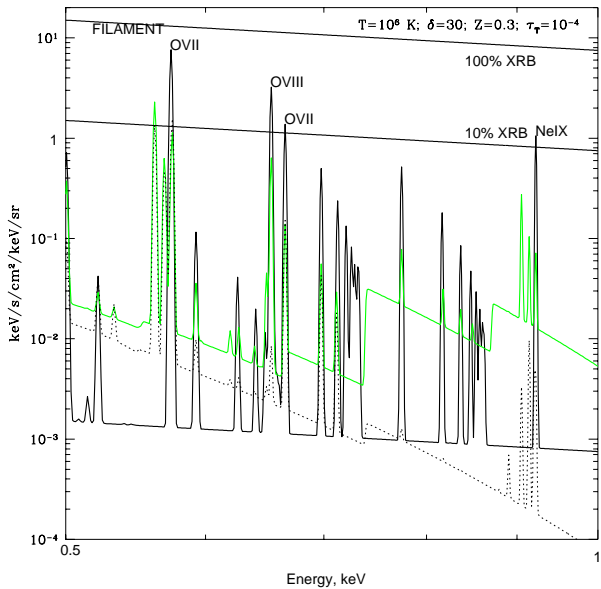}
\caption{The X--ray spectrum of warm gas with an overdensity
of $\delta=30$, a temperature  of $10^6$K, a Thomson optical
depth of $10^{-4}$ and a metallicity of 30\% solar.  These conditions 
should be typical for a filamentary structure in the warm IGM. 
The dotted curve shows the emission spectrum due to collisional 
excitation and ionization. For the  light solid curve 
photoionization by the XRB was taken into account. The dark solid  
curve shows  Thomson plus resonantly scattered radiation. 
The upper solid line is the intensity of the XRB from 
the same region. The lower solid  line shows  10\% of the XRB 
intensity (the level to which discrete source may be removed). 
The spectra were convolved with a  Gaussian with a FWHM of 2 eV. 
\label{specf}}
\end{figure}

\begin{figure}
\plotone{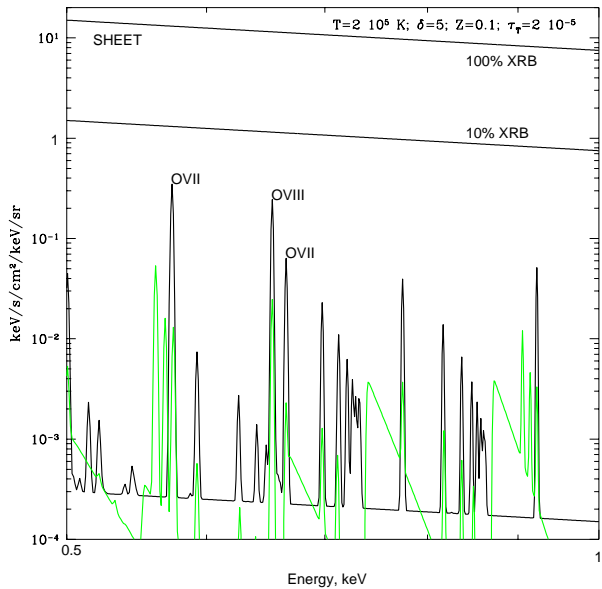}
\caption{The same as in Fig.1 but for warm gas with an overdensity of
$\delta=5$, a temperature   
of $2~10^5$K, a Thomson optical depth of $0.2 \times 10^{-4}$ and 
a metallicity of 10\% solar.
These conditions 
should be typical for a sheet-like structure in the warm IGM. 
The thermal emission from such a gas in pure collisional equilibrium 
is orders
of magnitude below the limits of the plot. 
\label{specs}}
\end{figure}

\subsection{Influence of the XRB on the thermal emission}
\begin{figure}
\plotone{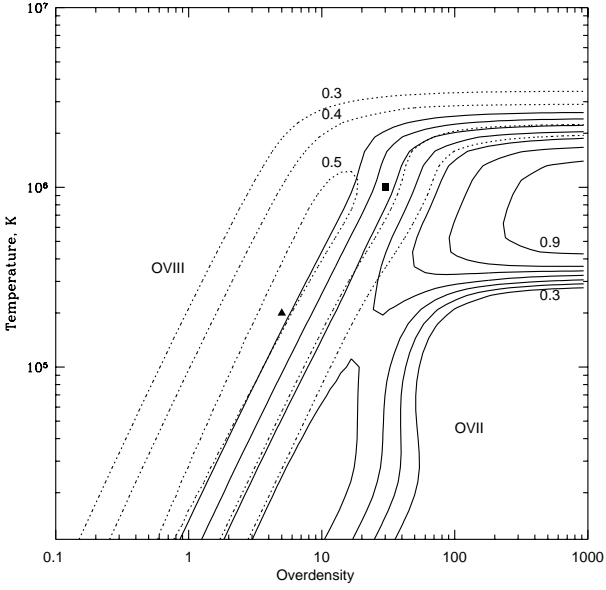}
\caption{Contours show the fraction of OVIII (dotted lines) and OVII (solid
lines) ions as a function of temperature and (over)density. Contours start
with 0.3 with 0.1 increment. Some of the contours are labeled. 
Ionization  equilibrium was calculated taking
into account photoionization by  XRB photons. Black square and triangle mark
the typical parameters of a filament and sheet respectively.
\label{o78}
}
\end{figure}

The IGM is exposed to  XRB photons. These photons change the
ionization balance of the IGM, producing ions at higher ionization
stage than  expected for pure collisional ionization at a given
temperature. The importance of photoionization depends on the
temperature and density of the gas.  The total ionization rate of a given
ion is 
\begin{eqnarray}        \label{ionrate}
n_e <\sigma v>_{\rm ion} + 4\pi \int I(E)\sigma_{\rm ph}(E) dE, 
\end{eqnarray} 
where $n_{e}$ is the electron density, and $<\sigma v>_{\rm ion}$ is
the collisional ionization rate (in $\cm^{-3}~\sec^{-1}$) which is a
function of temperature. The second term in
eq.(\ref{ionrate}) accounts for  photoionization, where $I(E)$ is the
background intensity ($\phot~\cm^{-2}~\sec^{-1}~\keV^{-1}~\sr^{-1}$) 
and $\sigma_{\rm ph}(E)$ is the photoionization cross section. We used the 
approximations for photoionization cross sections given by Verner \&
Yakovlev (1995) and Verner et al. (1996).  For simplicity we 
neglected ejection of multiple electrons which
may follow innershell ionization. For the other processes affecting the
ionization balance (i.e. collisional ionization and photo and
dielectronic recombinations) we used the values adopted in the MEKA
code. The characteristic time for photoionization of oxygen ions by
XRB photons, $t=\left ( 4\pi \int I(E)\sigma_{\rm ph}(E) dE\right
)^{-1}$ is  $\sim 3.5\times 10^9~\yr$  for \ovii\ and   $\sim
10^{10}~\yr$  for \oviii . This is somewhat shorter than the Hubble
time and  ionization equilibrium is approximately
established\footnote{For the high  abundances of \ovii\ and \oviii\ 
in which we are interested here  the recombinations time scales 
are of the same order.}.
At the densities and temperatures typical for the warm IGM the
oxygen is mainly in the form of He and H--like ions, as shown in
Fig.\ref{o78}. The symbols (dots for OVIII and circles for OVII ions
respectively) show the areas on the temperature/overdensity plot where
the fraction of He and H--like ions of oxygen is larger than 30\%.  \ovii\ 
and \oviii\ more or less trace the density temperature relation of the
warm IGM. A fraction larger than  30\% is expected for practically
the whole range of densities  and temperatures prevalent in the
warm IGM (see also Hellsten et al. 1998).

The change of the ionization balance due to photoionization 
affects the bound-free radiation,
recombination lines and strength of the lines excited by electron
collisions. The free-free emission does not change compared to the gas
at the same temperature in collisional equilibrium.
The corresponding thermal emission spectra (i.e. the sum of
the free-free, bound-free continuums, recombination and collisionally
excited lines) are shown by the grey lines in Fig.1,2 for our canonical  
filament and sheet, respectively. The change of the ionization 
state due to photoionization 
strongly enhances the X--ray emissivity of the gas. 
Photoabsorbed XRB photons are effectively converted into 
recombination radiation (in the form of bound-free radiation 
and recombination lines). The effect is especially strong when 
the temperature of the gas is low, e.g. around $10^5~K$. In this case 
the gas does hardly emit any X--rays if photoionization is neglected. 

\begin{figure}
\plotone{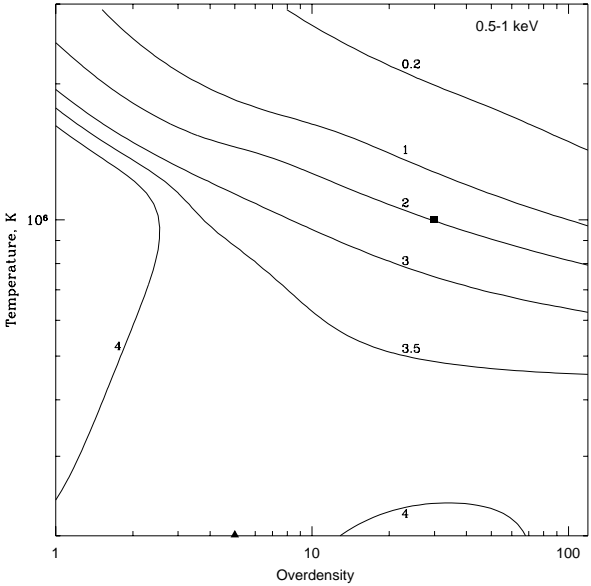}
\caption{Contour plot of the ratio of  scattered emission to 
thermal emission in the energy range 0.5--1 keV for typical  temperatures and
(over)densities. The solid square and triangle mark
the parameters of the canonical filament and sheet, defined in Section
2.1, respectively.
\label{frsce}
}
\end{figure}

\subsection{Resonant scattering {\it vs} local thermal emission}
Resonant scattering of XRB photons is even more important 
than the enhancement of the thermal emission 
of the gas due to photoionization. The emissivity of 
resonantly scattered radiation
($\cm^{-3}\sec^{-1}$) can be written as, 
\begin{eqnarray} \label{rsint}
\epsilon_{\rm rsc}(E_{\rm l}) =4\pi I(E_{\rm l}) n_i \sigma_0, 
\end{eqnarray} 
where $E_{\rm l}$ is the line
energy and  $n_{i}$  is number density of a given ion. Hereby
$\sigma_0=\frac{\pi e^2}{m_e c}f_{\rm ul}F$, where $f_{\rm ul}$ 
is the oscillator strength, $e$ is the electron charge, $m_e$ is 
the electron mass, $c$ is the speed of
light, and $F=4.14\times 10^{-18}$ is the conversion factor from Hz to
keV. The number density of a given ion is 
\begin{eqnarray} \label{ni}
n_i=  n_{_{\rm H}} Z_\odot Z f_{_{\rm X}},
\end{eqnarray}
where $n_{_{\rm H}}$ is the hydrogen density in the warm IGM, $Z_\odot$ is
the solar abundance of a given element relative to hydrogen,
$Z$ is the abundance of the element relative to solar abundance and 
$f_{_{\rm X}}$ is the fraction of atoms in a given ionization state. 
 
The resonantly scattered flux in a given line is proportional to the
number density of the ion. It is convenient to express its intensity
in terms of the intensity of the Thomson scattered continuum, which is
proportional to the density of the electrons. The emissivity of the
Thomson scattered continuum is given by,  
\begin{eqnarray}        \label{tsint}
\epsilon_{\rm tsc}(E_{\rm l})=4\pi I(E_{\rm l}) n_e \sigma_{_{\rm T}}, 
\end{eqnarray} 
where $n_e$ is the electron density and $\sigma_{_{\rm T}}$ is the Thomson
cross section. The total scattered spectrum (i.e. the sum of resonantly
and Thomson scattered radiation) is shown in Fig.1,2 by the thick
solid curves\footnote{Here we use the list of the strong resonant lines 
compiled by Verner ($http://www.pa.uky.edu/~verner/atom.html$).}. 

\begin{table}
\caption{Equivalent width of the resonant lines of H and He--like ions
relative to the Thomson scattered continuum
assuming unity  for the fraction of the
ionization state and solar abundances.
The abundances are taken
from Feldman (1992).}
\begin{center}\begin{tabular}{lcrr}\hline \\
Ion & Abundance ($Z_\odot$)& Energy (keV)& $E_{_{\rm EW}}$ (keV) \\
\\
\hline
 CV   &     3.98E-04 &    0.31 &     46.6 \\
 CVI  &              &    0.37 &     27.5 \\
 NVI  &     1.00E-04 &    0.43 &     11.7 \\
 NVII &              &    0.50 &     6.9\\
 OVII &     8.51E-04 &    0.57 &     99.6\\
 OVIII &              &    0.65 &     58.9\\
 NeIX &     1.29E-04 &    0.92 &     15.1\\
 NeX &              &     1.02 &     8.9\\
 MgXI &     3.80E-05 &     1.34 &     4.4\\
 MgXII &              &     1.47 &     2.6\\
 SiXIII &     3.55E-05 &     1.85 &     4.1\\
 SiXIV &              &     2.00 &     2.4\\
 SXV &       1.62E-05 &     2.45 &     1.9\\
 SXVI &              &     2.62 &     1.1\\
 FeXXV &     3.24E-05 &     6.67 &     3.8\\
 FeXXVI &              &     6.97 &     2.2\\
\hline
\end{tabular}
\end{center}
\end{table}

The ratio of resonantly and  Thomson
scattered emissivity (i.e. the equivalent width) is,
\begin{eqnarray}        \label{ew}
E_{_{\rm EW}} \sim \frac{\pi e^2}{m_ec}f_{\rm ul}\frac{Z_\odot
Z f_{_{\rm X}}}{\sigma_{_{\rm T}}}
\end{eqnarray} 

This expression is of course only valid for an optically
thin medium.  The equivalent widths for He and H--like ions
are given in Table 1  assuming unity for the fraction of the
ionization state and solar abundances. Along an isoelectronic sequence
(e.g. for He--like ions of heavy elements) the oscillator strength for a
given transition is approximately constant. Not surprisingly  the
resonant  transitions of H and He-like oxygen (\ovii\ and \oviii )
are particularly strong due to the high oxygen abundance.\footnote{For
neutral gas an analogous relation exists between the 
Thomson scattered continuum and intensity of the fluorescent lines
(e.g. Vainshtein, Sunyaev 1980). Resonant scattering and
photoionization  have a comparable effective cross sections, 
but an additional branching ratio (radiative decay vs autoionization
-- Auger effect) enters the expression for the equivalent width
in the case of fluorescent lines.  This branching ratio is small for 
light elements (e.g. $\sim$0.005 for the oxygen 1s-2p transition) 
and reaches $\sim$0.3 for iron. Because of this the equivalent width 
of the fluorescent lines  from neutral gas is always much lower (except for
the iron line at 6.4 keV) than in  
the case of resonant scattering for H- and He-like ions.}   

\begin{figure}
\plotone{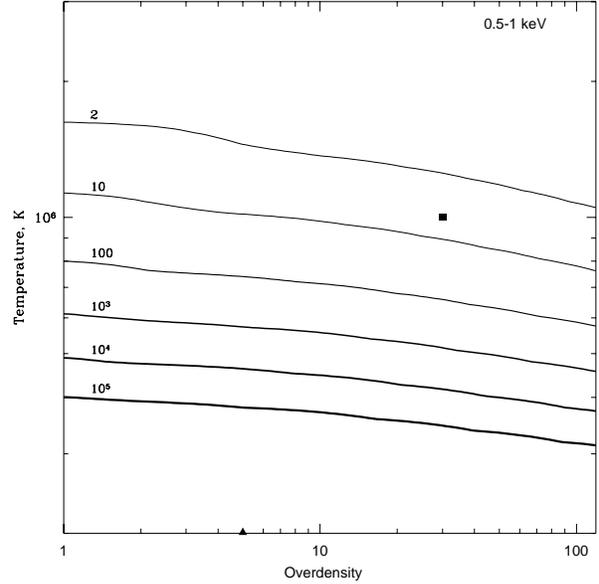}
\caption{Contour plot of the ratio of the total emissivity (including
scattered radiation) of gas photoionized by the  XRB 
and the thermal emission from  gas at the same density and temperature 
in pure collisional equilibrium. The emissivities are integrated over the
0.5--1 keV band. The solid square and triangle mark
the parameters of the canonical filament and sheet, defined in Section
2.1, respectively.
\label{pnop}
}
\end{figure}

We now  compare the emissivity of the gas due to scattering  and
that due to thermal emission (taking into account photoionization by
XRB). The ratio depends strongly on the temperature and density of the
gas as shown in Fig.\ref{frsce}. Both scattered and thermal emission
were integrated over the energy range 0.5--1 keV. At low densities and low
temperatures the ratio is about 3--4. For  these parameters 
photoionization strongly dominates over collisional ionization. As a result 
the ratio of the scattered and thermal
emission (the latter being dominated by recombination radiation) is
proportional to the 
factor $\frac{\sum n_i I(E_{\rm l}) \sigma_0}{\sum n_i \int
I(E)\sigma_{\rm ph}(E) dE}$, where the summation in the denominator is
over all ions and in the numerator over all ions and lines. For strong
individual lines this ratio is about 3--4. For our canonical filament
the 0.5--1 keV emissivity of the gas due to resonant scattering
exceeds the thermal emission of the photoionized gas by a factor of
about 2. 
In Fig.\ref{pnop} we compare the emissivity (including scattered radiation) of
the gas photoionized by the XRB to the pure thermal emission of the gas 
if it were in collisional ionization equilibrium. The 
ratio of the emissivities (integrated over the 0.5--1 keV
energy band) is shown as a contour plot. For our canonical filament 
this ratio is about 5--10 while for the sheet it is more than
$10^5$. We conclude that 
estimates for the detectability of the warm IGM which neglect either 
scattering of the soft X-ray background and/or photoionization 
of the IGM by  the soft X-ray background are  overly pessimistic. 

\begin{figure}
\plotone{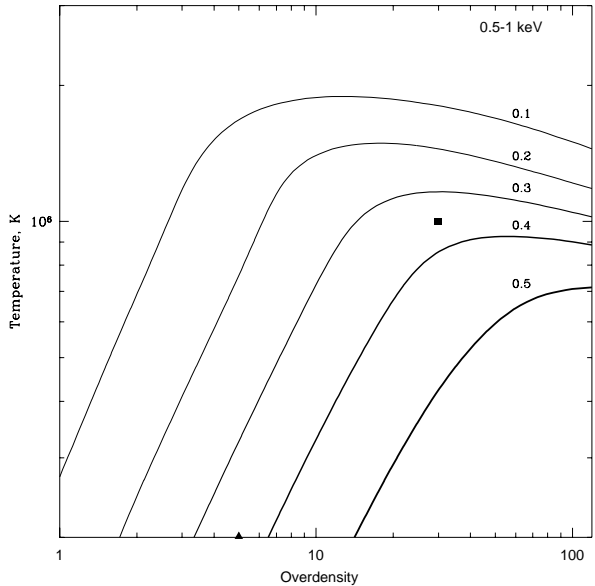}
\caption{
Contour plot of the fraction of the total emissivity in the 0.5--1
keV band which is due to   resonantly scattered radiation in the OVII
line at 0.57 keV. The solid square and triangle mark
the parameters of the canonical filament and sheet, defined in Section
2.1, respectively.
\label{frl}
}
\end{figure}

Finally,  we compare the emissivity due to resonant scattering 
in  the OVII line which is likely to be the strongest line at the
relevant temperature and densities,
to  the total integrated  emissivity in the energy range 0.5--1 keV 
(Figure \ref{frl}). For a typical filament about 30 percent of the 
total flux is  emitted in  the resonant line of  \ovii\ . 

\subsection{The contribution of resonant scattering to the diffuse
X-ray background} 
 
The intensity of  a background
scattered by a medium with 
uniform density is given by a formula similar to the Gunn-Peterson 
relation (Gunn and
Peterson, 1965, Shapiro and Bahcall 1980, Aldcroft et al., 1994).
In an Einstein-de-Sitter Universe this takes the form, 
\begin{eqnarray}        \label{xrbs}
\frac{I_{\rm rsc}(E_0)}{I_{_{\rm XRB}}(E_0)}=\frac{c}{H_0} 
Z Z_\odot f_{_{\rm X}} 
n_0\frac{\sigma_0}{E_0} (1+z)^{0.5} .
\end{eqnarray} 
Here $I_{\rm rsc}(E_0)$ is the intensity of the scattered background at the
observed  energy $E_0$ ($E_0<E_{l}$).  
Equation (\ref{xrbs}) assumes that temperature, abundance and 
ionization state of the gas do not evolve with redshift and 
that the XRB is due to distant sources. For the $1s^2$--$1s2p^1P$
transition of He--like oxygen the above
gives approximately,
\begin{eqnarray}        \label{xrbsrat}
\frac{I_{\rm rsc}(E_0)}{I_{_{\rm XRB}}(E_0)}\sim 0.04 \left(
\frac{Z}{0.3}\right )  \left( \frac{f_{_{\rm OVII}}}{0.3}\right ).
\end{eqnarray}
The  exact contribution of resonantly  scattered photons to the XRB 
is difficult to assess and will depend on the  detailed 
density, temperature and metal distribution of the IGM. From equation 
(\ref{xrbsrat}) we can, however, infer that the contribution should be at the
percent level for energies below the strong resonance lines of 
oxygen. 

From equation (\ref{xrbsrat}) we
can also see that the warm IGM is  optically thin in the strong \ovii\
resonance line up to  an overdensity  of at least  30 even if the
metallicity is  high and  the \ovii\ fraction is large.  For other
lines the optical  depth will be generally smaller.

\begin{figure}
\plotone{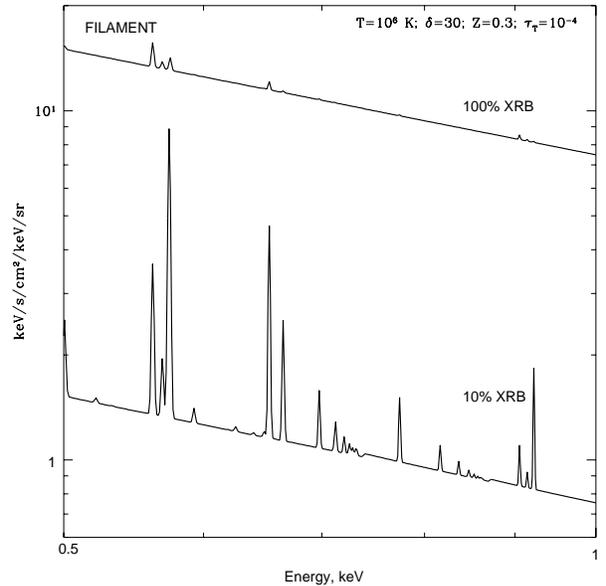}
\caption{The upper curve shows the total spectrum of a region of the sky
containing a filament with  Thomson optical depth of
$10^{-4}$. This spectrum is a sum of the thermal emission of the filament,
scattered XRB and emission of the background sources, part of which is
either absorbed or scattered by the filament. Note that resonance scattering
does not change the XRB flux and only emission lines due to the thermal
emission of the filament are visible in this spectrum.
The lower curve shows the spectrum of the same region if 90\%
percent of the background is resolved into compact sources and
removed. Note that the most prominent OVII line at 0.57 keV becomes
visible only in this spectrum. In the hypothetical case when all XRB is
resolved and removed the residual spectrum should be equal to the sum of the
thermal emission of the filament and scattered XRB emission. These two
components are shown in Fig. 1.
\label{netspec}}
\end{figure}

\begin{figure}
\plotone{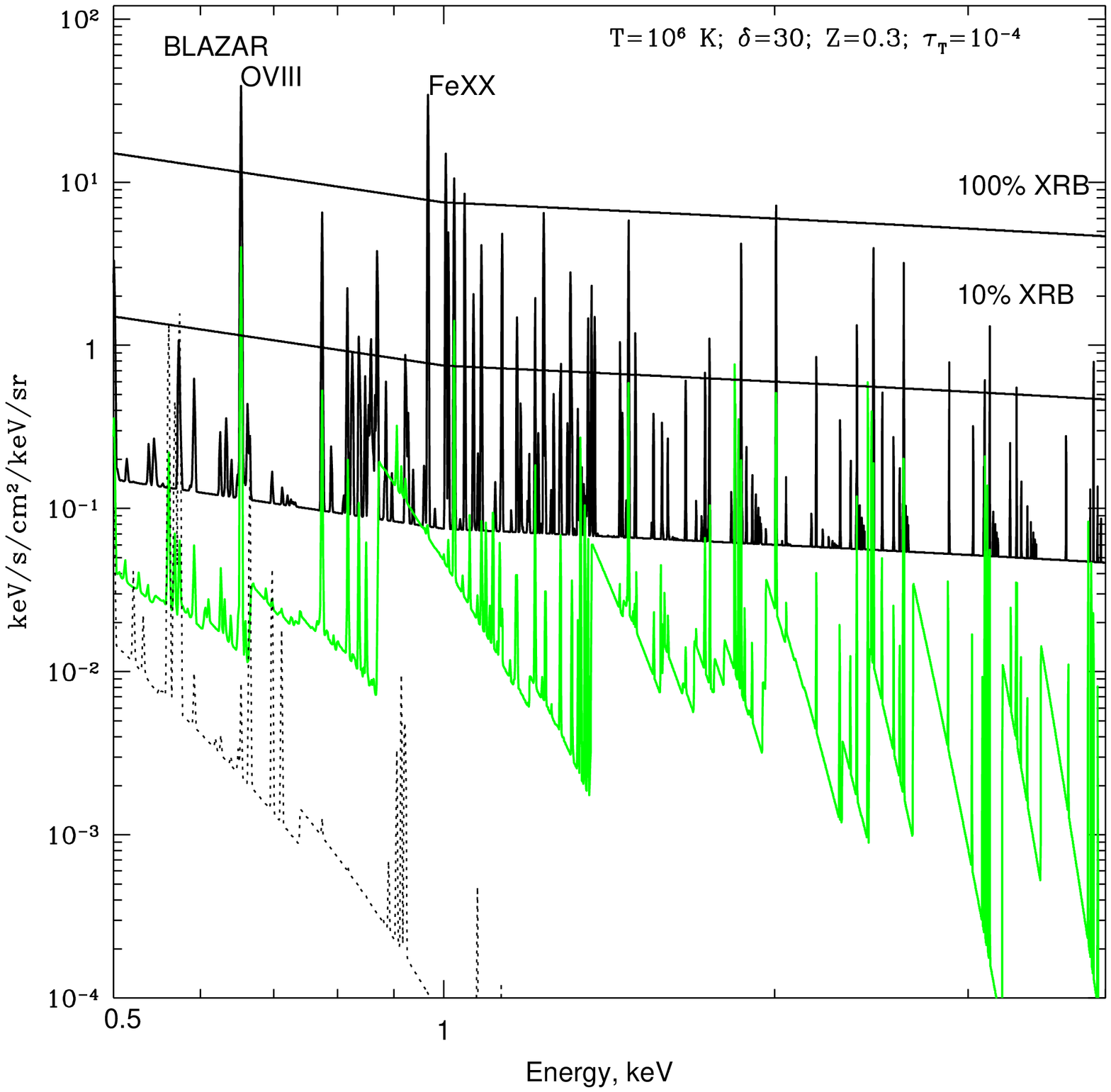}
\caption{
 Same as Figure 1  but assuming that the photoionizing continuum is
factor of 100 more intense than XRB. This figure illustrates the case
of a filament illuminated by a nearby bright quasar. 
\label{blazspec}}
\end{figure}

\section{Detectability of filamentary structures in the warm IGM} 

\subsection{Detecting resonance scattering from diffuse gas in emission}

Note that neither the number nor the energy of photons  change during resonant
scattering. Resonant scattering would not change the XRB flux if the 
XRB  were  completely homogeneous and isotropic\footnote{
To a smaller extent this is also true 
for the photoabsorbed photons of the XRB which are reemitted as 
recombination photons.}.  However, 
the XRB is emitted by discrete sources  and resonance scattering   
converts photons emitted by  compact 
sources into a diffuse background. 

The resonantly scattered emission 
will be detectable  in images in which discrete sources 
making  up a significant fraction of the total background 
have  been removed. 
This is demonstrated in Fig.\ref{netspec} which shows the  spectrum 
of our canonical filament (taking into account 
the  XRB)  and the case were  
90 percent of the background has been removed. ROSAT has e.g.  
resolved  70-80\% of the XRB in the energy range 0.5--2 keV
(Hasinger et al., 1998). At the lower end of this range the situation
is somewhat unclear mainly due to the emission of our own  Galaxy. 

The spectral resolution of X-ray instruments is dramatically
increasing. XMM and Chandra will have the first instruments with high
spectral resolution (using gratings) but only for bright compact
sources. With projected missions like Constellation-X and XEUS 
imaging with a spectral resolution of about 2 eV will become possible. 
The large ratio  of resonantly scattered emission to  the Thomson 
scattered  continuum and the local thermal emission makes the search for 
the spectral feature of resonantly  scattered radiation very worthwhile.
Detecting the warm IGM by  photons  scattered 
in the \ovii\ or \oviii\ resonant transitions 
has another big  advantage. Filaments and sheets contain a 
large number of faint galaxy clusters and galaxy groups.
The emission from the dense hot gas in these clusters and groups 
generally dominates the thermal emission from filaments. 
However, in this dense and hot gas  oxygen is generally  
completely stripped from electrons and no resonant scattering 
will occur in these regions.  The resonantly  scattered radiation 
will  thus be a good tracer of the diffuse gas in filaments.  
The only contamination should be due to resonant oxygen scattering 
by the warm gas which may be contained in the galaxies within the
filaments. 

The highest signal to noise ratio can be achived if 
\ovii\ or \oviii\ absorption lines are observed in the spectrum of a
very bright compact 
source located behind the filament as first suggested by Shapiro and Bahcall
(1980). The longest exposures (deep surveys) however are usually collected
for fields without strong X--ray sources. 
For such fields the detection of \ovii\ or \oviii\ resonant lines in
emission in the residual background is favorable compared to the
detection of these lines in absorption using the combined spectrum of all
resolved background sources in the field. The number of line photons
in the residual background 
is approximately equal to the number of line photons absorbed from the
spectra of all resolved sources. The intensity of the residual
background is, however, lower if most of the background is
resolved. This results in a higher signal-to-noise ratio for the
residual background. It means that in order for such a method to be
useful the particle and other ``non cosmological'' detector backgrounds have
to be low compared to the combined intensity of the resolved sources
in the field.   

\subsection{Detecting filaments with upcoming satellite missions}

\begin{table*}
\caption{Parameters of different X-ray missions. The number quoted in
the table are approximate and should only be used for 
rough estimates. For XEUS 
the  parameters are quoted for the Narrow and Wide  field
imagers, respectively.  The total number of 
(resonantly scattered) photons in the OVII line 
which will be collected from a filament filling the field of view (FOV) of the
telescope is proportional to the product of the effective area and
the size of the FOV. This product (normalized to
XMM) is given in a row labeled as $\frac{S}{S_{XMM}}$. The signal-to-noise
ratio (assuming pure statistical errors mainly due to background
photons from unresolved sources and Galactic emission in a band pass set  
by the energy resolution of the telescope) is given in the last row of 
the table.}
\begin{center}\begin{tabular}{lcccc}\hline \\
&XMM (EPIC pn)& Chandra&XEUS&Constellation-X\\
\\
\hline
Area@0.5 keV, $cm^2$  & 1300 & 260 &  30000& 5000\\
FOV, $\arcm$ &$25\times 25$& $18\times 18$& ($1\times 1$;$5\times 5$)
& $3\times 3$\\
$\Delta E$, eV &60&100&(1;50)&2eV\\
$\frac{S}{S_{XMM}}$ &$1$&$0.1$&($0.04$;$0.92$)&$0.06$\\
$\frac{S/N}{S/N_{XMM}}$ &$1$&$0.25$&($1.1$;$1.1$)&$1.3$\\
\hline
\end{tabular}
\end{center}
\end{table*}

In an Einstein de Sitter Universe 
our canonical filament of  comoving size ($16 h^{-1}\Mpc \times 3 h^{-1}\Mpc$ 
has angular extent $(183'\times 34',50'\times 9',31'\times 6')$      
at $z = (0.1, 0.5,1)$, respectively. 
In the case of a filament,
completely filling the FOV the expected number of counts in the oxygen
line is proportional to the product of the effective area and 
the FOV (solid angle) of the telescope.
This value is largest for XMM and  is comparable for the 
Wide field imager of XEUS. 
Estimates of the effective area, FOV
and energy resolution are summarized  in Table 2.
For a nearby ($z \la  0.1$) filament with Thomson optical depth 
of $10^{-4}$ XMM (EPIC pn detector)
will detect about 200 counts in the OVII line 
in a  $10^5$ s exposure. 
For Chandra the expected number of counts is about an order of magnitude
lower. 

The  total particle background in a 60 eV band pass
(the resolution of XMM at 0.5 keV)   will be at the level
of $\sim$50--60 counts for a $10^5$ s exposure. The residual
unresolved extragalactic background (assuming that 90\% will be
resolved) will produce $\sim$ 1000 counts.  
The biggest problem is probably Galactic emission.
Galactic emission  is usually modeled as
a combination of two thermal (i.e. thermal emission of optically thin
plasma) components with temperatures $\sim$0.15 and 0.05 keV
respectively (e.g. Miyaji et al., 1998). A  redshift of 0.1 
should, however, be sufficient to distinguish  oxygen lines 
produced by  a filament from  oxygen lines produced in the Galaxy. 
Galactic  emission  will produce several thousand counts in a 
60 eV wide band pass. 
For a $10^5$ second  XMM exposure the signal-to-noise ratio will thus 
be  of order a  few. Similar signal-to-noise ratios can be achieved 
for the OVIII line at 0.65 keV.  Such signal-to-noise ratios
should be sufficient for a meaningful cross-correlation with the redshift 
distribution of galaxies in the same field.  
Longer exposure times,  filaments with  larger Thomson optical 
depth, mapping of larger areas of sky or illumination 
by a nearby bright X-ray source as described in the next section 
will  be necessary  to increase the signal-to-noise ratio. 

The  anticipated  FOV of Constellation X and of the Narrow field 
imager of XEUS is rather small (see Table 2). The  total number of
photons detected in the OVII and OVIII lines for a  
filament filling the whole FOV will therefore be an order of 
magnitude lower for these missions than  for XMM for a  
similar exposure time(except for the Wide field imager of XEUS, which
will detect a similar number of photons).  
The signal-to-noise ratio (assuming pure statistical errors due to 
residual unresolved extragalatic and Galactic background within 
the band pass set by the energy resolution of the telescope) 
will nevertheless be similar due to the better energy resolution 
of XEUS and Constellation X. For
an efficient study of the diffuse emission of filaments a larger 
FOV would be very important. If it were  possible to increase the
FOV to e.g. $5'\times 5'$ (below 1 keV) for  high energy
resolution detectors like those of Constellation X and XEUS then these
missions should be able to detect   resonantly scattered
emission of filaments along any line of sight during a 50ks
exposure.

\section{Illumination by a nearby bright source}

Close to a  bright source the emission due to resonant scattering 
is obviously  enhanced both relative to 
the local emission and  relative to the  background. 
Bright sources may have soft X-ray luminosities of $10^{46} \erg \sec^{-1}$ 
or larger and exceed the soft X-ray background by a factor $(r/r_{0})^{-2}$
out to a radius $r_{0} \sim 30 (L/10^{46} \erg \sec^{-1})^{0.5} \Mpc$
and the signal-to-noise ratio will be enhanced by the same factor.      
Particular interesting is the illumination of 
strong filaments close to rich galaxy 
clusters by either the X-ray emission  from the core of the
cluster or by a bright AGN contained in the cluster. 
This is illustrated in Fig.\ref{blazspec}, where we 
show the intrinsic  and scattered emission for the same 
parameters as in Fig. 1, but with a factor 100 larger 
intensity of the illuminating continuum. For simplicity we assumed the
same shape for the quasar spectrum as for the XRB. The increased 
X-ray flux also affects the ionization state of the gas.  
The relevant quantity is the ionization parameter which is
proportional to the ionizing flux.  
The regions in  Fig.\ref{o78} will e.g. shift linearly  
to the right with increasing flux (at least approximately 
for the moderate densities  and temperatures where collisional 
ionization is not important). 
This  shifts the dominant ionization state from \ovii\ to
\oviii .  The strongest line in Fig.\ref{blazspec} is indeed that of
\oviii\ at 0.65 keV. The local thermal emission is also strongly
enhanced (compare the dotted to the grey and thick solid curves in
Fig.\ref{blazspec}).  Very close to bright sources  oxygen may 
be even completely stripped due to photoionization.  
The fraction of H--like ions is then inversely
proportional to the blazar flux while the number of
scattered photons (e.g. in the H-like 1s-2p transition) 
is proportional to the blazar flux and the density of 
H-like ions. The resonantly scattered flux will then not depend on 
the illuminating flux anymore and will instead be proportional to 
the recombination rate.  The number of resonantly scattered
photons will still be larger than the number of 
recombination line photons by a factor of few. On the other hand the
Thomson scattered flux is simply proportional to the intensity of
illuminating radiation. The relative importance of the Thomson
scattered continuum is thus strongly enhanced close to a strong 
ionizing source.

The number of bright AGN which are available 
for an illumination of filaments is much larger than naively 
inferred from the AGN luminosity function. A considerable fraction of 
bright AGN are blazars with strongly beamed radiation. If these
typically have an  opening angle of 5 degrees there will be 
about 2000 objects which point away of us for any observed blazar. 

Note that in the case of illumination by individual bright sources the 
assumption of an isotropic phase function for the resonant scattering 
which we made so far is not valid. The phase function 
will depend on the particular transition. For the $1s^2$--$1s2p^1P$ 
transitions of He--like ions the phase function will e.g. be the 
familiar dipole (see e.g. Chandrasekhar 1960) and the scattered
radiation will be polarized. 

The illumination of filaments by an AGN 
may also constrain  the lifetime of these objects. 
If the active phase of AGN lasts for about $10^7$years 
then the observed size of the illuminated region will be 
of the order of 3 Mpc (depending on the angle of the 
beam to the line of sight). 
Of course the observed flux in the
scattered lines depends on a number of factors (density,
temperature, blazar flux, heavy elements abundance etc.). 
 Various structures in the surface
brightness which may appear as a result of scattering of beamed
radiation are considered by Gilfanov, Sunyaev \&  Churazov (1987) and
Wise \& Sarazin 1990. 

The time, needed to establish photoionization equilibrium 
is inversely proportional to the photoionization rate and is equal to
$\sim 10^9 \left (\frac{D}{20~\Mpc} \right )^{2}~\left 
(\frac{L}{10^{47}\erg\sec^{-1}}\right )^{-1}~\yr$. Thus if the blazar 
was active for a shorter period of time or the beam direction varies
(e.g. due to precession of the jet) then the influence of the 
photoionization by the blazar emission will be lower accordingly.
The analysis of gas illuminated by a nearby sources is therefore
not  straightforward. It will, however, be  easier 
to interpret for beamed sources than  for isotropic sources.  

\section{Conclusions}

We have demonstrated that for the temperatures and densities 
prevalent in the the filamentary and sheet-like structures of 
the warm intergalactic medium illumination by  XRB photons
significantly increases the X--ray emissivity of the medium. 
The strongest contribution to the X--ray emissivity is due to
resonant scattering of XRB photons. This resonantly scattered 
emission can be detected in images from  which  a significant fraction 
of the XRB due to compact sources has  been removed.

Resonant lines of He and H-like oxygen are most promising
for detecting filamentary structures in the warm IGM. 
The fraction of oxygen in the ionization states \ovii\ or \oviii\ 
is  larger than 30 percent practically for the whole range 
of densities  and temperatures expected in the warm IGM. 

Estimates for the detectability of filaments in the  
warm IGM  which take only local thermal emission into account 
may have been overly pessimistic. XMM should be able to detect 
the radiation resonantly  scattered by diffuse gas in a filament 
with Thomson optical 
depth of $10^{-4}$ and  metallicity 0.3 solar at $z\sim 0.1$ with 
signal-to-noise of a few.

For  filaments close to massive cluster the signal will  be 
enhanced  due to illumination from the cluster core
and/or AGN contained in the cluster. The most promising sources 
emitting high X-ray intensities are beamed blazars which should 
have a high space density.  For AGN and especially blazars 
constraints on the duration of the active phase may also be 
obtainable.

\section{Acknowledgements} 
We thank the referee K. Yamashita for helpful comments.
MH gratefully acknowledges the hospitality of the Institute 
for Theoretical Physics  Santa Barbara where this research 
was completed. This research was supported in part by the National Science
Foundation under Grant No. Phy94-07194.

\label{lastpage}

\end{document}